# A New Approach to Component Testing

Dr. Horst Brinkmeyer
Ingenieurbüro Brinkmeyer, Leibnizstr.11, 70806 Kornwestheim
horst.brinkmeyer@ibb-kwh.de

## Abstract

*Carefully tested electric/electronic components are a requirement for effective hardware-in-the-loop tests and vehicle tests in automotive industry. A new method for definition and execution of component tests is described. The most important advantage of this method is independance from the test stand. It therefore offers the oppportunity to build up knowledge over a long period of time and the ability to share this knowledge with different partners.*

## 1. Introduction

Complexity of electric/electronic systems in automotive environment has increased significantly in the last ten to fifteen years. The components are designed by engineers in supplier companies, but the layout of the entire electric/electronic system is done by the car manufacturers themselves. This job has become more difficult because of the increasing complexity, reduced development time and reduced budgets. So effectivity and reliability are very important topics in this field of engineering. According to the V-modell the task consists of layout and testing of the entire electric/electronic system. In this paper only testing will be observed.

For the OEM testing normally consists of two stages:
- hardware-in-the-loop tests (HIL) and
- tests within the vehicle itself.

Either of these test methods is extremly costly and the goal is to find out, weather there are difficulties in the interaction of the components that haven't been forseen during design and weather there are unexpected conditions in the entire system. Important for effective engineering in this field are components that behave according to their requirements. As the component tests normally are executed by the supplier, the OEM has to rely on this; but the task for the supplier often is difficult as well: The written requirements for the components are nomally incomplete and the knowledge of the supplier about possible difficulties may be incomplete as well, if the developpement of similar components has been carried out by other suppliers. In this situation OEM and supplier are in danger to repeat mistakes and therefore the components delivered for HIL tests and vehicle tests may not have the expected quality. As a result of this, time and money is lost or, even worse, the quality of the car affected.

To avoid this, a method is needed, to preserve the knowledge about requirements of components, including bugs, that have occured in the past. On one hand there is of course the specification of the component but the requirements also have to be tested. So there is a need for test cases that are specfied in a way, so that a high percentage of them can be reused in order to perserve the experience for future projects.

## 2. Requirements for component tests

Component tests today normally are defined in a specific test environment i.e. for a determinated test hard- and software. In this case knowledge between the OEM and the supplier can only be transferred, if both of them define component tests in the same environment. In most cases the OEM doesn't define component tests at all and so knowledge about test cases is lost from one project to another. In order to change this situation, the following requirements for component tests are derived:

- **Component tests have to be independent from the test environment**
  This means, stimuli and expected outputs for the device unter test have to be defined only according to the requirements of the component itself and not according to the properties of the test stand.
- **The tool for test definition must be easy to use**
- **The test definitions must be easy to read**
- **A tool is needed for automatic generation of code, that can be interpreted by any test stand**
- **Interpreters for the generated code are needed for those test stands, that are going to be used for component tests**

## 3. The tool chain for definition of component tests

A lot different tools exist for definition of tests; usage of most of them normally is limited to especially qualified and trained personell. In order to allow usage of the tool chain to all involved engineers without specific training, we choose Excel as input tool for the test definition. Three different types of Excel sheets are used to describe the test sequence:
- Signal definition sheet
- Test definition sheet
- Status definition sheet

In the signal definition sheet all input and output signals of the device under test (DUT) are defined as well as the status of these signals before starting the test itself.

The tests themselves are defined in test definition sheets. In each test only a certain part of the specification is testet; according to this part of the specification, only those signals are mentioned in this sheet that describe the behaviour of the DUT under this point of view. For each test step status are assigned to one or more signals. Status assigned to signals describe either the stimulus to be applied to an input signal or an expected output or output range in the case of output signals. The following table shows a simple example:

| test step | Δt | IGN_ST | DS_FL | DS_FR | NIGHT | INT_ILL | remarks |
|---|---|---|---|---|---|---|---|
| 0 | 0,5 | Off | Closed | Closed | 0 | Lo | day: no interior |
| 1 | 0,5 | | Open | | | Lo | illumination, if |
| 2 | 0,5 | | Closed | Open | | Lo | doors are open |
| 3 | 0,5 | | | Closed | | Lo | |
| 4 | 0,5 | | Open | | 1 | Ho | night: interior |
| 5 | 0,5 | | Closed | | | Lo | illumination on, |
| 6 | 0,5 | | | Open | | Ho | if doors are open |
| 7 | 280 | | | | | Ho | |
| 8 | 25 | | | | | Lo | illumination |
| 9 | 0,5 | | | Closed | | Lo | off after 300s |

The behaviour of the signal INT_ILL (interiour illumination) is described as a function of the signals IGN_ST (ignition status), DS_FL (door switch front left), DS_FR (door switch front right) and the bit NIGHT, comming from a light sensor. If the bit NIGHT is active, the interior illumination is lit for a maximum duration of 300s, if one of the doors is open, what is indicated by an "Open" status of the door switch.

The expressions **Off, Open**, **Closed**, **0**, **1**, Lo and Ho are status, which are defined in the so called status table. The status table for the example may look like this:

| status | method | attribut | var (x) | nom | min | max | D 1 | D 2 | D 3 |
|---|---|---|---|---|---|---|---|---|---|
| Off | put_can | data | | 0001B | | | | | |
| Open | put_r | r | | 0 | | | 0,5 | 1 | 2 |
| Closed | put_r | r | | INF | | | INF | 5000 | 5000 |
| 0 | put_can | data | | 0B | | | | | |
| 1 | put_can | data | | 1B | | | | | |
| Lo | get_u | u | UBATT | 0 | 0 | 0,3 | | | |
| Ho | get_u | u | UBATT | 1 | 0,7 | 1,1 | | | |

Each status used in the signal table or test table is defined in the status table. First of all a method is mentioned that is used to apply the status. For example the status "Lo" or "Ho", defining high or low voltage at a pin, is carried out by the the method "get_u". This method mesasures the voltage at the specified signal pin and compares it to the limits. For example the status "Ho" is valid, if the measured voltage is between 0,7*Ubatt and 1,1*Ubatt, where Ubatt is the supply voltage of the DUT.

The tests defined by this type of Excel sheets now have to be transformed to a form that can be interprted easily by a test stand. As file type we have chosen the xml-Format. Besides header, step numbers etc. the most important content of this file is given by many signal statements, each of them followed by a method statement. For example the following code represents checking of the "Ho"-status for the signal INT_ILL:

```
<signal name="int_ill">
    <get_u  u_max="(1.1*ubatt)" u_min="(0.7*ubatt)" />
</signal>
```

## 4. Execution of component tests

The generated xml-File (test script) can be used for test execution at test stands equipped with the affiliated interpreter. Besides the test script, the test stand needs information about its own ressources and in which way these ressources can be connected to the DUT. Ressources in this context are described by the methods that are supported by them and the valid range for all parameters.

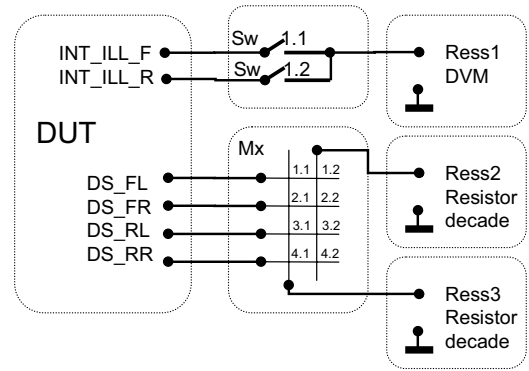

In our example there are three resources, one DVM and two resistor decades, that can be connected to the DUT. The DVM supports the method "get_u", the resistor decades the method "put_r". The following tables show the representation of this test circuit for the test stand:

| Ress. | Method | Attribut | Min | Max | Unit |
|---|---|---|---|---|---|
| Ress1 | get_u | u | -60 | 60 | V |
| Ress2 | get_r | r | 0 | 1,00E+06 | Ω |
| Ress3 | get_r | r | 0 | 2,00E+05 | Ω |

| | INT_ILL_F | INT_ILL_R | DS_FL | DS_FR | DS_RL | DS_RR |
|---|---|---|---|---|---|---|
| Ress1 | Sw1.1 | Sw1.2 | | | | |
| Ress2 | | | Mx1.2 | Mx2.2 | Mx3.2 | Mx4.2 |
| Ress3 | | | Mx1.1 | Mx2.1 | Mx3.1 | Mx4.1 |

For each method to be carried out, the test stand searches an approriate ressource, that can be connected to the signal pin. If this is not possible an error message is generated.

## 5. Status of the project

The described method for component testing was developed for DaimlerChrysler and successfully applied to two ECUs of the next S-class. At the moment different projects are carried out for suppliers.